\documentclass[letterpaper,twocolumn,prl,aps,superscriptaddress,amsmath,amssymb,floatfix]{revtex4-1}
\usepackage{mathptmx}
\usepackage[latin9]{inputenc}
\usepackage{color}
\usepackage{amsmath}
\usepackage{amssymb}
\usepackage{graphicx}
\usepackage{esint}
\usepackage[unicode=true,
 bookmarks=true,bookmarksnumbered=false,bookmarksopen=false,
 breaklinks=false,pdfborder={0 0 1},backref=false,colorlinks=true]
 {hyperref}
\hypersetup{
 linkcolor=magenta,urlcolor=blue,citecolor=blue,pdfstartview={FitH},hyperfootnotes=false}

\makeatletter

\usepackage{textcomp}
\usepackage{epstopdf}

\pdfpageheight\paperheight
\pdfpagewidth\paperwidth

\@ifundefined{textcolor}{}{%
 \definecolor{BLACK}{gray}{0}
 \definecolor{WHITE}{gray}{1}
 \definecolor{RED}{rgb}{1,0,0}
 \definecolor{GREEN}{rgb}{0,1,0}
 \definecolor{BLUE}{rgb}{0,0,1}
 \definecolor{CYAN}{cmyk}{1,0,0,0}
 \definecolor{MAGENTA}{cmyk}{0,1,0,0}
 \definecolor{YELLOW}{cmyk}{0,0,1,0}
}

\usepackage{xcolor}\usepackage{soul}
\newcommand{\bra}[1]{\ensuremath{\left\langle#1\right|}}
\newcommand{\ket}[1]{\ensuremath{\left|#1\right\rangle}}

\definecolor{blue}{rgb}{0,0,1}
\definecolor{red}{rgb}{1,0,0}
\definecolor{green}{rgb}{0,1,0}

\usepackage{soul}

\makeatother

\begin{document}
\title{Hardware-Efficient Bosonic Module for Entangling Superconducting Quantum Processors via Optical Networks }

\author{Jia-Hua~Zou}
\affiliation{Laboratory of Quantum Information, University of Science and Technology of China, Hefei 230026, China}
\affiliation{Anhui Province Key Laboratory of Quantum Network, University of Science and Technology of China, Hefei 230026, China}

\author{Weizhou~Cai}
\affiliation{Laboratory of Quantum Information, University of Science and Technology of China, Hefei 230026, China}
\affiliation{Anhui Province Key Laboratory of Quantum Network, University of Science and Technology of China, Hefei 230026, China}

\author{Jia-Qi~Wang}
\affiliation{Laboratory of Quantum Information, University of Science and Technology of China, Hefei 230026, China}
\affiliation{Anhui Province Key Laboratory of Quantum Network, University of Science and Technology of China, Hefei 230026, China}

\author{Zheng-Xu~Zhu}
\affiliation{Laboratory of Quantum Information, University of Science and Technology of China, Hefei 230026, China}
\affiliation{Anhui Province Key Laboratory of Quantum Network, University of Science and Technology of China, Hefei 230026, China}

\author{Qing-Xuan Jie}
\affiliation{Laboratory of Quantum Information, University of Science and Technology of China, Hefei 230026, China}
\affiliation{Anhui Province Key Laboratory of Quantum Network, University of Science and Technology of China, Hefei 230026, China}

\author{Xin-Biao~Xu}
\affiliation{Laboratory of Quantum Information, University of Science and Technology of China, Hefei 230026, China}
\affiliation{Anhui Province Key Laboratory of Quantum Network, University of Science and Technology of China, Hefei 230026, China}

\author{Weiting~Wang}
\affiliation{Center for Quantum Information, Institute for Interdisciplinary Information Sciences, Tsinghua University, Beijing 100084, China}

\author{Guang-Can~Guo}
\affiliation{Laboratory of Quantum Information, University of Science and Technology of China, Hefei 230026, China}
\affiliation{Anhui Province Key Laboratory of Quantum Network, University of Science and Technology of China, Hefei 230026, China}
\affiliation{Hefei National Laboratory, Hefei 230088, China}

\author{Luyan~Sun}
\email{luyansun@tsinghua.edu.cn}
\affiliation{Center for Quantum Information, Institute for Interdisciplinary Information Sciences, Tsinghua University, Beijing 100084, China}
\affiliation{Hefei National Laboratory, Hefei 230088, China}

\author{Chang-Ling~Zou}
\email{clzou321@ustc.edu.cn}
\affiliation{Laboratory of Quantum Information, University of Science and Technology of China, Hefei 230026, China}
\affiliation{Anhui Province Key Laboratory of Quantum Network, University of Science and Technology of China, Hefei 230026, China}
\affiliation{CAS Center For Excellence in Quantum Information and Quantum Physics, University of Science and Technology of China, Hefei, Anhui 230026, China}
\affiliation{Hefei National Laboratory, Hefei 230088, China}

\date{\today}

\begin{abstract}
Scaling superconducting quantum processors beyond single dilution refrigerators requires efficient optical interconnects, yet integrating microwave-to-optical (M2O) transducers poses challenges due to frequency mismatches and qubit decoherence. We propose a modular architecture using SNAIL-based parametric coupling to interface Brillouin M2O transducers with long-lived 3D cavities, while maintaining plug-and-play compatibility. Through numerical simulations incorporating realistic noises, including laser heating, propagation losses, and detection inefficiency, we demonstrate raw entangled bit fidelities of $F \sim 0.8$ at kHz-level rates over 30~km using the Duan-Lukin-Cirac-Zoller (DLCZ) protocol. Implementing asymmetric entanglement pumping tailored to amplitude damping errors, we achieve purified fidelities $F \sim 0.94$ at 0.2\,kHz rates. Our cavity-based approach outperforms transmon schemes, providing a practical pathway for distributed superconducting quantum computing.
\end{abstract}
\maketitle

\noindent \textit{\textbf{Introduction}.-}
Over the past two decades, superconducting quantum processors have achieved remarkable milestones, from demonstrating quantum computational advantage~\cite{Arute2019,Wu2021} to the protection of quantum information through quantum error correction (QEC)~\cite{Ofek2016,Chen2023,Ni2023,Acharya2025,Brock2025}, establishing the superconducting chips as a leading platform in quantum technologies. However, scaling beyond single dilution refrigerators faces fundamental constraints in the cooling power and limited space. Distributed quantum computing (DQC) architectures promise further scalability for eventual fault-tolerant quantum computing~\cite{Barral2024, Jiang2007, Caleffi2024}, and are also attractive for blind quantum computing~\cite{Barz2012}, clock synchronization~\cite{Komar2014}, distributed quantum sensing~\cite{Gottesman2012}, and fundamental tests of of quantum mechanics~\cite{Hensen2015}. Superconducting DQC requiring efficient quantum interconnects over room-temperature low-loss optical fiber networks, demands high-performance microwave-to-optical (M2O) quantum transducers. Consequently, substantial experimental and theoretical efforts over the past decade have pursued efficient and low-noise M2O conversion~\cite{Han2021,VanThiel2025,Arnold2025,Warner2025}.

Despite significant progress in demonstrating M2O transduction and microwave-optical entanglement~\cite{Zhong2020,Sahu2023,Meesala2024}, most efforts focus on direct photon conversion or entangling single microwave modes with optical fields. While optically heralded entanglement of superconducting systems was proposed conceptually~\cite{Krastanov2021}, a complete and experimentally feasible architecture delivering high-rate and high-fidelity remote entanglement under realistic conditions remains absent. Practical deployment faces critical challenges, including the thermal noises and finite conversion efficiency of M2O transducers, the frequency mismatches between the superconducting devices and transducers, and the operation errors and the decoherence of the quantum registers. Furthermore, long-distance entanglement distribution protocols are typically heralding-based~\cite{Briegel1998, Duan2001, Simon2003, Duan2004, Barrett2005, Childress2006, Sangouard2011, Azuma2012, Azuma2012b}, the long photon flying time, combined with this non-deterministic nature, requires a long qubit lifetime to support tasks involving multiple remote e-bits, such as entanglement purification, before decoherence occurs. These constraints collectively make the e-bits generated in heralding-based protocols are scarcely useful for practical applications.



In this Letter, we present a hardware-efficient bosonic module architecture that provides a complete blueprint for scalable superconducting quantum networking. Our modular design addresses the core challenges through a SNAIL-based parametric interface enabling plug-and-play transducer integration with flexible frequency matching and dynamic noise isolation; long-lived 3D cavity memories supporting high-rate entanglement generation and subsequent purification operations; realistic performance optimization accounting for dominant noise sources and operation errors. Through comprehensive numerical simulations, we demonstrate that remote entanglement with fidelities $F \sim 0.8$ at kilohertz rates (raw) or $F \sim 0.94$ at hundreds of hertz (purified) is achievable using near-term technology, providing a practical pathway toward distributed superconducting quantum computing.

\begin{figure*}
\begin{centering}
\includegraphics[width=1\linewidth]{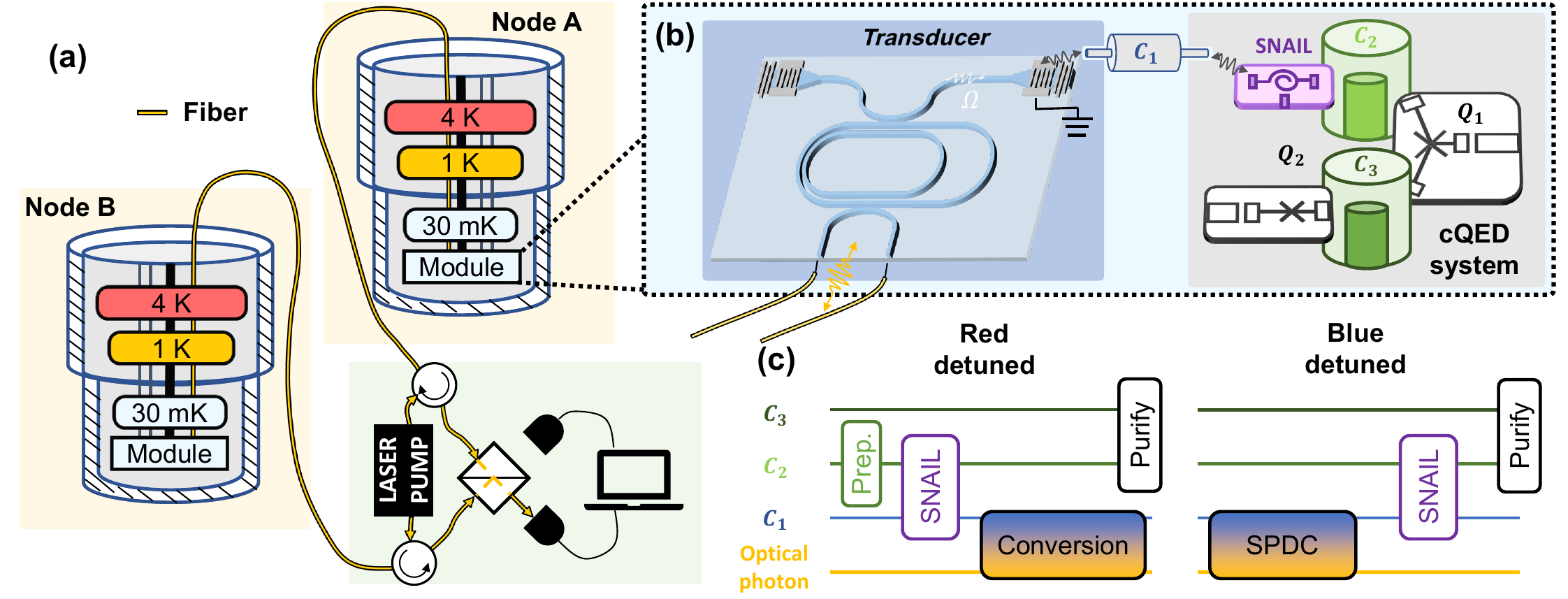}
\end{centering}
\caption{(a) Schematic diagram of the superconducting quantum network incorporating the proposed module.
A laser pumps the transducers at both distant nodes, where the microwave photons are converted into optical photons and transmitted back for coincidence measurement.
By post-selecting single-click detection events, the remote modules become entangled.
(b) Schematic of the module: each unit consists of one transducer and one corresponding circuit quantum electrodynamics (cQED) system.
(c) Timing diagram illustrating the establishment of microwave-optical photon entanglement via both direct conversion and spontaneous parametric down-conversion (SPDC).}
\label{fig1}
\end{figure*}

\noindent \textit{\textbf{Hardware-efficient bosonic module}.-} Figure~\ref{fig1}(a) sketches the optical fiber link for generating entanglement between superconducting bosonic module in separate dilution refrigerator. In each node, a bosonic module is placed in 30\,mK environment. The module comprises two integrated subsystems, as illustrated in Fig.~\ref{fig1}(b): a M2O transducer and a circuit quantum electrodynamics (cQED) processor~\cite{Cai2024,Zhou2025}. A high-quality aluminum coaxial cable cavity ($C_1$) is bonded to the transducer chip, and is capacitively coupled to a superconducting nonlinear asymmetric inductive element (SNAIL)~\cite{Frattini2017APL,Hua2025}. The cQEC processor consists of two long-lived 3D microwave cavities ($C_2$, $C_3$), serving as quantum registers, coupled to two ancillary transmon qubits ($Q_1$, $Q_2$) that enable universal quantum gate operations between $C_2$ and $C_3$ through numerically optimized pulses~\cite{Heeres2017,Khaneja2005,Krastanov2015,Heeres2015,Eickbusch2022} and high-fidelity readout.

We employ a M2O transducer chip based on stimulated Brillouin scattering to realize the efficient conversion between traveling phonons and optical phonons~\cite{Yang2024,Yang2025}. This platform is based on the thin-film lithium niobate on sapphire, which supports simultaneous high-quality phononic and phononic cavities modes, and also the efficient unidirectional inter-digital transducer (UDT) based on piezoelectric coupling for microwave photon-phonon conversion. The Brillouin M2O benefits the module from two aspects: on one hand, the suspension-free devices ensures excellent thermal dissipation, which is critical for suppressing the laser pump heating effect. On the other hand, the potential optical stray light that be deleterious to the superconducting devices are isolated from the superconducting components.

The bosonic module holds practical advantages for realizing high-rate remote superconducting entanglement from several aspects: (i) the SNAIL provides parametrically-driven nonlinear couplers~\cite{Miano2022SNAIL}, allows flexible frequency matching between the M2O transducer and the resonator $C_2$, and also the impedance matching efficient conversion between the phonons on the transducer chip and microwave excitations in the $C_2$ cavity. (ii) the bosonic modes provide quantum storage with long coherence time~\cite{Milul2023,Ganjam2024} essential for probabilistic heralding entanglement generation protocols. Additionally, the high-dimensional Hilbert space also offers potential for advanced QEC encoding schemes~\cite{Cai2021}. (iii) The dual-cavity architecture with ancillary qubits enables sophisticated quantum operations including entanglement purification~\cite{Dur1999,Jiang2010,Krastanov2021} and potentially fault-tolerant control. (iv) Lastly, the SNAIL-based connection can be high-efficiency and interchangeable, thus enabling plug-and-play operations for excellent scalability~\cite{Mollenhauer2025}.


\begin{figure*}
\begin{centering}
\includegraphics[width=1\linewidth]{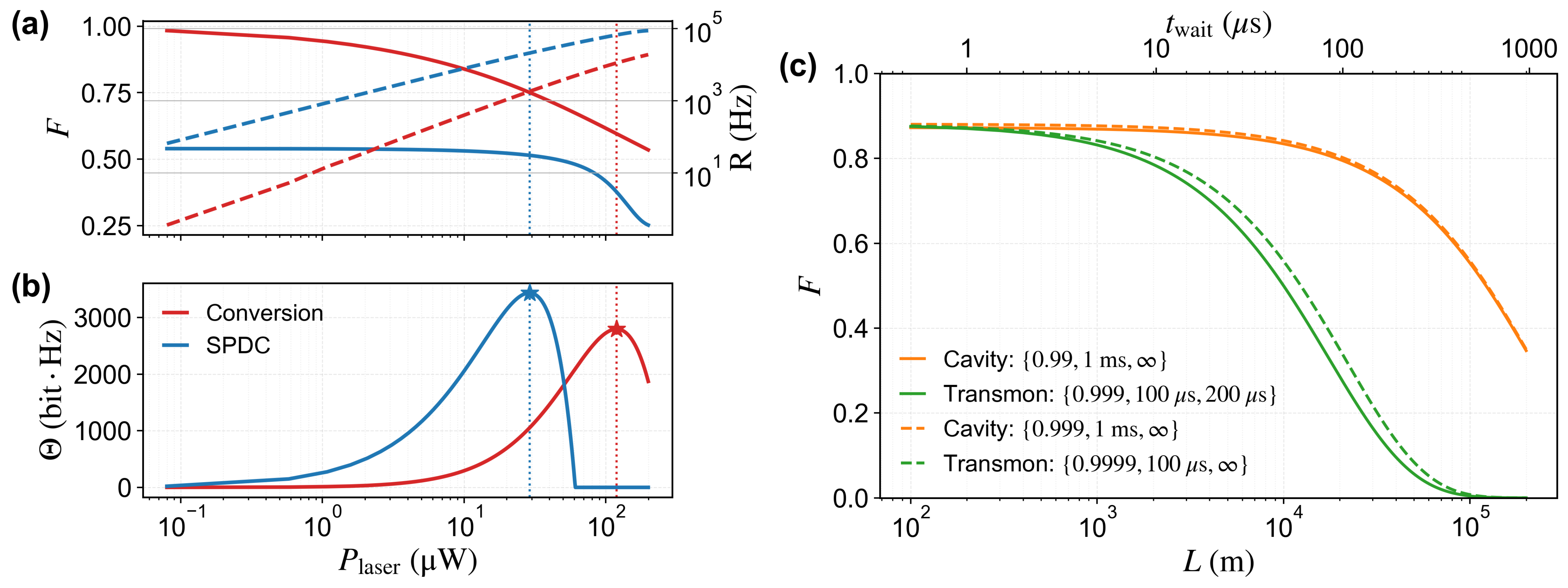}
\end{centering}
\caption{(a) Raw e-bit distribution performance in both conversion and SPDC schemes.
Solid lines represent fidelity $F$, and dashed lines correspond to the heralding rate $R$ (right $y$-axis).
For the conversion scheme, $P_e$ is optimized to ensure the highest fidelity $F$ at each pump power.
(b) Raw e-bit throughput: the highest-throughput points are marked with stars in corresponding colors.
(c) Comparison between transmon and cavity in entanglement distribution.
The parameters inside the curly braces represent the operation fidelity $F_{\mathrm{op}}$, energy relaxation time $T_1$, and pure dephasing time $T_{\phi}$, respectively.
The node distance $L$ is assumed to be 1~$\mathrm{km}$ in (a) and (b).}
\label{fig2}
\end{figure*}

\smallskip
\noindent \textit{\textbf{Single photon emission}.-} We focus on the heralded entanglement generation, as firstly conceptualized in Ref.~\cite{Krastanov2021}, with the imperfections introduced by the M2O conversion can be mitigated through subsequent entanglement purification. While direct quantum state transfer through the module is theoretically possible, binomial QEC code requires a total transfer efficiency $\eta>81\%$ to overcome dominant photon loss errors~\cite{Michael2016,Hu2019}. However, state-of-art M2O converters only promises the $\eta\leq10\%$~\cite{Yang2024}.

Figure~\ref{fig1}(c) illustrates the two schemes for generating entanglement between the cavity $C_2$ and optical photons, and the heralded entanglement between two modules can be realized by detecting coincident single photons emitted from both nodes via the DLCZ protocol~\cite{Duan2001}. In the red-detuned pump case, the optical pump stimulates the beam-splitter type phonon-photon interaction, leading to the coherent M2O conversion. Cavity $C_2$ is initialized in the Fock state $\ket{1}$, and the SNAIL partially releases the excitation to $C_1$ with a probability of $P_e$, and consequently converting the excitation into optical photon in the fiber through the transducer. This process is analogous to a solid-state single-photon emitter, producing the entangled state $\sqrt{1-P_e} \ket{1}_\mu\ket{0}_o + \sqrt{P_e} \ket{0}_\mu\ket{1}_o$, where $\ket{0}_{\mu(o)}$ ($\ket{1}_{\mu(o)}$) denotes the absence (presence) of a microwave (optical) photon.

In the blue-detuned pump case, the M2O transducer operates in the spontaneous parametric down-conversion (SPDC) mode, generating entangled optical photon-phonon pairs in the transducer chip. When the on-chip pump power $P_{\mathrm{laser}}$ is small, the resulting state can be approximated to first order as $\sqrt{1-P_e} \, \ket{0}_a\ket{0}_o + \sqrt{P_e} \, \ket{1}_a\ket{1}_o$, where $\ket{0}_a$ ($\ket{1}_a$) denotes the absence (presence) of an acoustic phonon, and $P_e$ is now the pair-generation probability determined by $P_{\mathrm{laser}}$. Simultaneously, as single photons are emitted into the optical fiber, the phonons are converted to microwave photons and swapped into $C_2$ via the SNAIL.


\smallskip
\noindent \textit{\textbf{Remote entanglement generation}.-} After a single-sided click event, assuming the photon detector has no photon-number-resolving capability, the physical qubits from both nodes A and B are expected to be projected into maximally entangled qubits (e-bits) $\ket{\psi^-}=\frac{1}{\sqrt{2}} (\ket{0}_{\mu,A}\ket{1}_{\mu,B} - \ket{1}_{\mu,A}\ket{0}_{\mu,B})$, serving as the resource states for further distributed quantum information processing over nodes. To evaluate the potential performance of the module determined by the imperfections of the interfaces and M2O transducers, we first consider an ideal cQED system and calculated the e-bit fidelity $F=\bra{\psi^-}\rho_{AB}\ket{\psi^-}$ and the corresponding heralding rate $R$ for an estimated density matrix ($\rho_{AB}$) of mode between two nodes. Realistic experimental imperfections are considered, including the transducer thermal noise due to the background temperature and pump laser heating~\cite{Xu2020,Xu2024,Fu2021}, multiple photon-phonon pair generation in SPDC mode, the microwave loss in the UDT for phonon waveguide-cable coupling $\eta_{\mathrm{UDT}}=0.6$, the optical loss in the optical waveguide-fiber coupling ($\eta_{\mathrm{opt}}=0.6$), propagation loss in long fiber ($0.2\,\mathrm{dB/km}$) with length $L$, and detector efficiency ($\eta_{\mathrm{QE}}=0.9$, no dark count~\cite{You2020}). The system operates in the pulse mode (duty cycle 1:1), with experimentally feasible parameters for Brillouin M2O transducer~\cite{Yang2024,Yang2025,Yang2025BSIT} including: a vacuum phonon-photon coupling cooperativity of $C_{\mathrm{Brillouin}}=8.8\times10^{-8}$, an optical (acoustic) intrinsic quality factor of $2\times10^6~(2\times10^4)$, and a waveguide extraction factor of $\zeta_{o(m)}=0.5~(0.9)$ for critically-(over-) coupled optical (acoustic) mode.

\begin{figure*}
    \begin{centering}
    \includegraphics[width=1\linewidth]{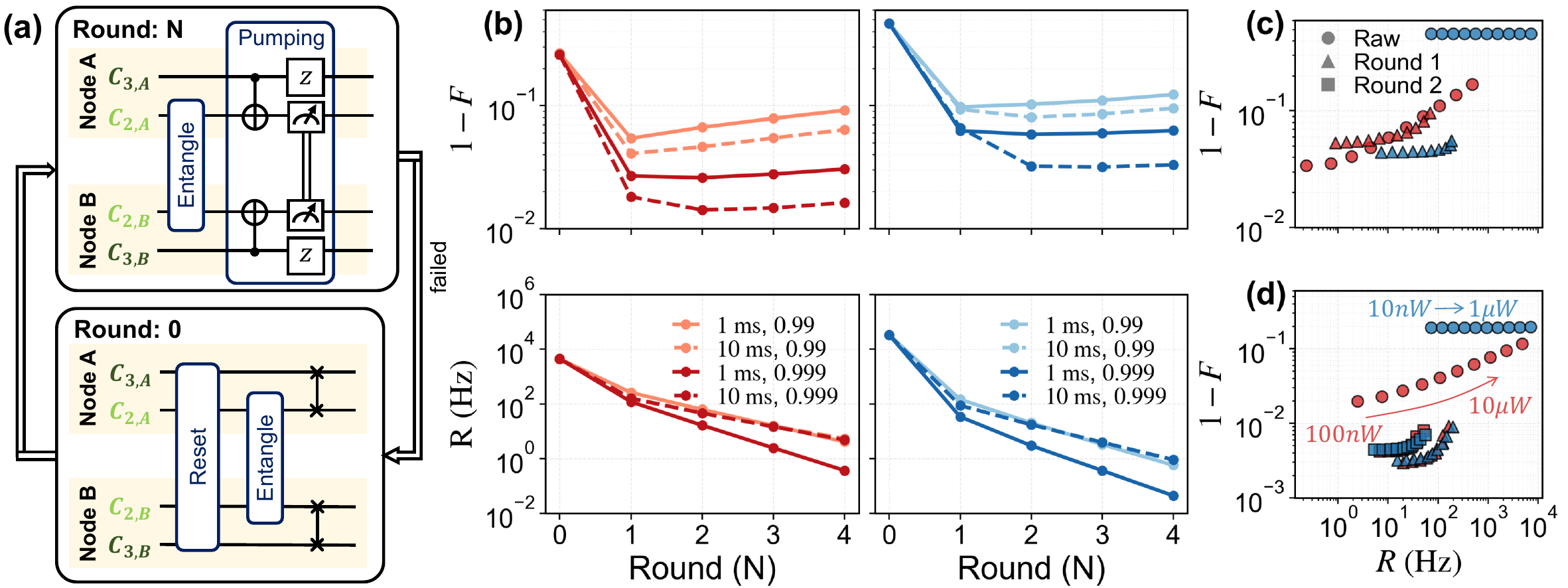}
    \end{centering}
    \caption{
(a) Schematic diagram of the entanglement pumping protocol. In round~0, a raw e-bit is prepared in $C_{3,A}$ and $C_{3,B}$. In each subsequent round, the circuit does not apply bilateral rotations to the e-bits before the CNOT gates~\cite{Bennett1996}; the control pair ($C_{3,A}$, $C_{3,B}$) is preserved only if the target pair ($C_{2,A}$, $C_{2,B}$) is measured in the state $\ket{1}_{\mu,A}\ket{1}_{\mu,B}$. The protocol restarts from round~0 if pumping fails.
(b) Entanglement pumping performance using the conversion and SPDC schemes. Parameters in the legend denote $T_1$ and $F_{\mathrm{op}}$, respectively. Initial raw e-bits in both scheme are obtained with $P_{\mathrm{on\text{-}chip}}=5~\mathrm{\mu W}$ and $P_e=0.25$.
(c) Comparison between raw and purified e-bits after one round of purification, with $T_1=1~\mathrm{ms}$ and $F_{\mathrm{op}}=0.99$. The raw e-bits in round~0 are generated under different pump powers, and the corresponding purified results are plotted for comparison.
(d) Comparison between raw and purified e-bits after up to two rounds of purification. The cQED system is idealized with $\eta_{\mathrm{UDT}}=0.9$, $T_1=10~\mathrm{ms}$, and $F_{\mathrm{op}}=0.999$.
Similarly, the raw e-bits in round~0 are obtained by varying $P_{\mathrm{laser}}$, as indicated by the colored arrows, and their purification outcomes are evaluated.
The $y$-axes in (b)--(d) show infidelity $1-F$ on a logarithmic scale for clarity.
}
    \label{fig3}
\end{figure*}

As shown in Fig.~\ref{fig2}(a),  $F$ reduces and $R$ increases with increasing pump power $P_{\mathrm{laser}}$ for both conversion and SPDC schemes. It is expected that increased laser power will stimulate the coherent Brillouin conversion process for high conversion efficiency, but also increases the device temperature for higher thermal excitations. Comparing the conversion and SPDC schemes, the SPDC has lower fidelity due to extra insertion losses in the phonon-microwave arm, which could not be mitigated by the heralding scheme, and thus the heralding rate is higher.

As shown in Fig.~\ref{fig2}(b), to further benchmark our raw e-bits, we introduce the the e-bit throughput $\Theta=R\times E_N(\rho_{AB})$ to evaluate the the upper bound of Bell states that can be distilled per second, with $E_N(\rho_{AB})>0$ being the logarithmic negativity of e-bits~\cite{Vidal2002}. The throughput of the conversion scheme saturates at around $100~\mathrm{\mu W}$ and the SPDC mode saturates at a lower pump power of around $20~\mathrm{\mu W}$. The pump power at which the throughput saturates indicates that further increasing the pump power would lead to a significant degradation of entanglement, and even result in purely false heralding events (as observed in SPDC when the power exceeds $50~\mathrm{\mu W}$). Therefore, it gives us a practical upper limit for the operating pump power of the transducer. Below this threshold, one can flexibly choose between higher fidelity and faster heralding rate by tuning the pump power.

Next, we consider the imperfections of a realistic cQED system, taking into account the energy relaxation time ($T_1$) and pure dephasing time ($T_\phi$), and also the quantum operation imperfections described by the general depolarization channel. Figure~\ref{fig2}(c) compares the performance of cavity-based and transmon-based (replace $C_2$ by a two-level qubit in the module) register qubits for generating e-bits across different node distances, corresponds to a waiting time ($t_\mathrm{wait}\sim$) in each e-bit generation cycle, with all systems operating in the conversion mode at $P_\mathrm{laser}=5\,\mathrm{\mu W}$. With experimentally feasible parameters of a transmon and a cavity~\cite{Chen2025Robust}, cavity outperforms transmon for distances $L>500\,\mathrm{m}$ owing to their long coherence times, although transmon with higher gate operation fidelity.
Even for scenarios with one order-of-magnitude improvement in operational error and negligible dephasing ($T_\phi = \infty$), the fidelities are remained almost unchanged, indicating that raw e-bit generation is fundamentally restricted by energy relaxation rather than local operational fidelity.

\smallskip
\noindent \textit{Entanglement pumping.-} With a raw e-bit fidelity lower than 0.8 at a rate $R\sim 1\,\mathrm{kHz}$ is achievable at a $t_{\mathrm{wait}}$ at sub-ms scale (or equivalently a distance beyond $10\,\mathrm{km}$), we next demonstrate how entanglement pumping~\cite{Dur1999,Jiang2010,Zhou2025} protocols can boost e-bit fidelity for potential scalable distributed quantum information processing. The scheme is illustrated in Fig.~\ref{fig3}(a), the raw e-bits on cavities $C_2$ are repetitively prepared and consumed to enhance the e-bit fidelity in cavities $C_3$, employing the local CNOT gates between the cavities followed by joint measurement on $C_2$ in the computational basis. If both measurements in Round $N$ yield $\ket{1}$ (success), the e-bit is retained for Round $N+1$, otherwise the states are discarded and the system resets to Round 0. Since the amplitude damping errors dominates in our bosonic modules, we employ asymmetric EP protocol, which allows us to purify e-bits even with a raw $F<0.5$, in contrasts to the DEJMPS scheme~\cite{Deutsch1996} designed for symmetric depolarization errors. It worth noting that a heralding time of successive raw e-bits generation $t_{\mathrm{wait}}\sim 1/R\ll T_1$ and high performance multiple cavities necessitate the employment of the bosonic module in potential realizations of entanglement pumping.

Figure~\ref{fig3}(b) presents the results of EPs for both conversion (red curves) and SPDC (blue curves) schemes with $L=50\,\mathrm{m}$, where the evolutions of infidelities ($1-F$) and the corresponding rate $R$ accounting the success probabilities upto to $N=4$ rounds are evaluated. In all cases, lower operation error and longer $T_1$ show improved e-bit fidelity after EP, while the generation rate is insensitive to those parameters. By repeating the EP, the $R$ shows exponential decreasing with $N$ as expected, while the fidelities showing saturation behaviors which is eventually reaches a balance due to the imperfections introduced by the operation errors and the accumulated error during the waiting times.

In Fig.~\ref{fig3}(c), a comprehensive comparison of bosonic module EP performances across different operating modes and pump powers is presented, revealing the basic trade-off relation between the fidelity and rate. For the Round 0 EP (raw e-bits), $F\sim 0.6$ with a $R$ exceed $10\,\mathrm{kHz}$ for SPDC mode, and $F>0.8$ with $R=1\,\mathrm{kHz}$ can be achieved with conversion. One pump round can boost $F\approx0.94$ with a high rate of $200\,\mathrm{Hz}$ for both SPDC and conversion modes, while the second pump round will degrade both $F$ and $R$. In Fig.~\ref{fig3}(d), the potential performances with improved $F_{\mathrm{op}}=0.999$ is shown. It is interesting to observe that conversion and SPDC modes show very similar performance given a 10-times low pump power in SPDC.

\smallskip
\noindent \textit{Conclusion.-}
We have proposed and analyzed a hardware-efficient bosonic module architecture for interconnecting superconducting quantum processors over optical networks. The bosonic module addresses two practical experimental challenges for scalable implementation of distributed quantum tasks: the plug-and-play M2O transducer compatibility and the capability in high-rate and high-fidelity e-bit preparation. Utilizing the Brillouin M2O transducer, SNAIL tunable coupler, and long-lived bosonic cavities, internode high-quality e-bit generation at 0.2\,kHz heralding rates with fidelities $F\sim0.94$, or kHz-level heralding rates to distribute raw e-bit ($F\sim0.8$) entanglement over 30\,km, are predicted with feasible experimental parameters, considering realistic (dominant) imperfections and noises are considered. All bosonic module architecture are compatible with the bosonic QEC and fault-tolerant control techniques, allowing the further exploration of remote e-bits with improved fidelity-rate trade-offs. Our results suggest that many distributed quantum protocols are accessible by combing advanced superconducting quantum circuits with state-of-the-art but imperfect M2O transducer components, providing a practical pathway toward scalable distributed quantum information processing.


\bigskip
\begin{acknowledgments}
This work was funded by the the Innovation Program for Quantum Science and Technology (Grant Nos.~2024ZD0301500 and 2021ZD0300200), National Natural Science Foundation of China (Grant Nos.~92265210, 92265108, 123B2068, 92165209, 92365301, 12474498, 11925404, 12374361, and 12293053). We also acknowledge the support from the Fundamental Research Funds for the Central Universities and USTC Research Funds of the Double First-Class Initiative. The numerical calculations in this paper were performed on the supercomputing system in the Supercomputing Center of USTC, and this work was partially carried out at the USTC Center for Micro and Nanoscale Research and Fabrication.
\end{acknowledgments}

\end{document}